# An Ontology-based Model for Indexing and Retrieval

*Winfried Gödert*
Cologne University of Applied Sciences
Institute of Information Science
Phone: ++4922182753388
winfried.goedert@fh-koeln.de

13.12.2013

Starting from an unsolved problem of information retrieval this paper presents an ontology-based model for indexing and retrieval. The model combines the methods and experiences of cognitive-to-interpret indexing languages with the strengths and possibilities of formal knowledge representation. The core component of the model uses inferences along the paths of typed relations between the entities of a knowledge representation for enabling the determination of hit quantities in the context of retrieval processes. The entities are arranged in aspect-oriented facets to ensure a consistent hierarchical structure. The possible consequences for indexing and retrieval are discussed.

*1. Introduction*

Subject indexing is no longer regarded as an area with large potential for methodological innovations. In practical applications, a decline of elaborated methods of intellectual indexing to represent the document's content, its aboutness, can be observed. A continuing use of intellectual indexing efforts can only be seen within the context of social tagging, but on a comparatively low level of differentiation of the content to be represented.

There is a different situation in the field of information retrieval. Developments of methods and procedures like link-topological or probabilistic approaches extend the basic Boolean retrieval model to new components. This results in substantial refinements of the hit quantity formation and the hit list display which can be achieved algorithmically without human work.

Fewer developments can be reported for indexing methods. The interest in automated processes has led to methods for identification of text tokens by using combinations of linguistic and statistical approaches. The correspondence of the estimated tokens with terms, their conceptual meaning or thematic contexts decreases with the aforementioned order of the three concepts. Retrieval success is often restricted to a character-based matching between search terms and stored data. Methods for disambiguation of indexing results and high potential to form selective sets of hits, for example combinations of faceted conceptual orderings with syntactic indexing approaches, have been invented. Although well known, they have become at least part of the current information systems. The intellectual effort and time required may well be the cause. Specification of the indexing results is commonly replaced by combinations of linguistic and statistical methods for generating hit lists and their ranked presentation.

Are the crucial problems of subject indexing solved? A closer look will show that this is not true. It is necessary to identify the still existing problems. Additionally, one should get an impression whether a problem may be solved by further improvements on the retrieval side, or what problems can be expected to be solved only by efforts undertaken on the side of indexing.

An example illustrates one of the open questions. Let us consider the following two search interests

- We are looking for documents about the *migratory instinct of songbirds*,
- We are looking for documents about *songbirds with migratory instinct*.

What is different between the two questions with relevance for a semantic representation, an indexing task, and a retrieval process? In the first question *songbirds* and *migratory instinct* are treated topics with a mutual reference between them. In order to use the two concepts as index terms, they must be included as elements in an appropriate indexing language. For a precise search result both terms need to be indexed, and an identification of the relationship between them in the document is required to achieve a co-extensive content representation. Expressed in the language of the indexing methods, a form of *syntactic indexing* and the use of *syntax operators* is required for the retrieval process. The method of *coordinating indexing* complemented by *post-coordinating retrieval* cannot guarantee the desired precision.

For the second search interest, *migratory instinct* shall not be a treated topic, so it may not have been indexed for the documents to be found. In this case, *migratory instinct* is only a constraining property for selecting the appropriate species of songbirds. Therefore it is a constraint for selecting the appropriate entities of the indexing language used for processing the search query and for generating the hit list. The anchoring of such property may be done alone in the knowledge representation and must be made available for a search operation by a suitable design of the retrieval environment.

We can generalize the situation: We are looking for documents on topics that are specified by certain properties. These properties should only be used as selection criteria for choosing the appropriate index terms but cannot be indexed because they are not topics treated by the document.

This distinction is impossible in the context of classical indexing and retrieval models. The classical models are characterized by the following properties: Finding document sets depends on the presence of index terms. The structure of the indexing language used - usually a form of hierarchy or subordination - is used only in a small number of cases to provide collections of index terms for the search query. Tools for distinguishing our previously mentioned two questions are not offered. According to current knowledge there is no way to take advantage from methods of automatic text analysis with statistical or linguistic approaches for treating these problems. We look for a solution that combines methods of formal knowledge representation with indexing and retrieval techniques.

For this, we provide an extension of the classical document-based indexing and retrieval model, which can be characterized by a combination of properties:

- Indexing of documents is a statement about the *aboutness* of documents, i.e. an indexing term is only assigned if the corresponding concepts are treated issues within the context of the document;
- The knowledge representation contains "knowledge" in form of an indication of properties of the entities contained, i.e. in the knowledge representation specific connections between concepts are stored by specifying properties and relationships;
- Inferences about the underlying knowledge, the entities and the structure between them are used to form sets of hits, i.e. not only the indexed entities, but also the relationships between them are taken as basis to form hit quantities.

*2. An ontology-based model for indexing and retrieval*

The underlying basis of the considerations to be presented is illustrated in figure 1.

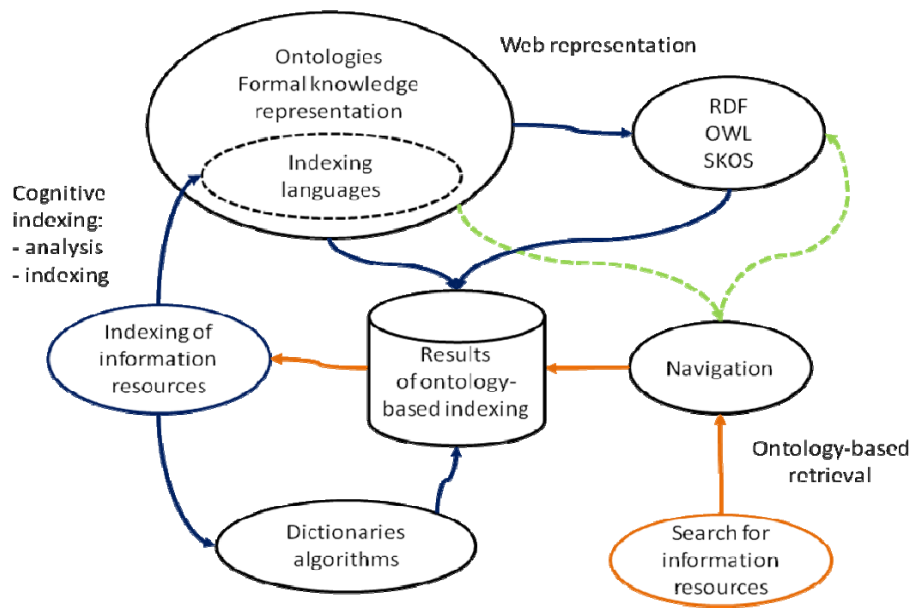

Fig. 1: Sketch of an ontology-based concept for indexing and retrieval

For a description of our model it is necessary to take a closer look on

- typed relations in knowledge representations as tools for representing different types of connections between concepts,
- forms of inferences along the relational paths usable for information retrieval.

Both aspects will be linked together for the goals of our discussion that was already started elsewhere (Gödert, 2010).

To simplify the way of speaking we state some prerequisites and make some agreements. The set of all structured terminological repositories (e. g. thesauri, classification systems), which can be used for indexing purposes shall be referred to as *knowledge representations*. They are distinguished in more detail, if required by the argumentation context. The knowledge representations consist of elements - called *entities* - and a structure that is determined by the relationships established between the elements. It is important to provide a compatibility between intellectual interpretation of the entities on the one side and their formal definitional characterizations by specifying attributes and relational properties on the other side. As structure-forming properties, the term relationships are seen in conjunction with aspect views to construct faceted structures.

*2.1 Inferences as aids of information retrieval*

The use of an index term for the formation of sets of hits can be seen as the simplest form of reasoning in the context of retrieval processes. Other examples of inference processes are offered by the use of the well-known tools of information retrieval, such as truncation or combining search terms by Boolean operators. Figure 2 shows an illustration of these situations.

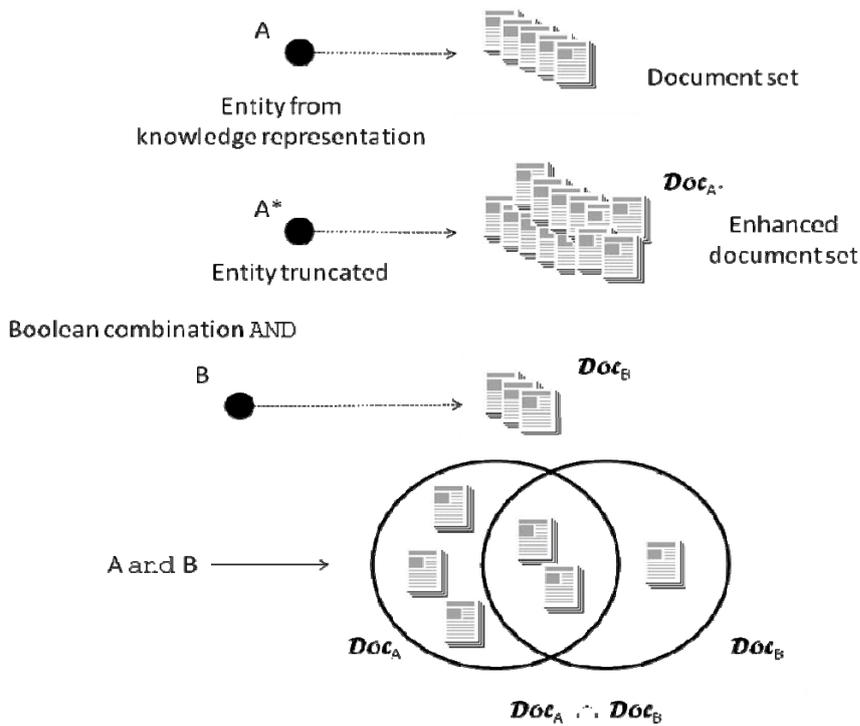

Fig. 2: Retrieval tools as inference operations

The operation on the set of entities creates a hit list by using the connection to the document collection that was established by the process of indexing. An extension of this form of reasoning consists of inferences along the relational paths that are created in the knowledge representation. Commonly known is the formation of thematic sections by use of hierarchical relationships, as is shown in figure 3 for a simple example.

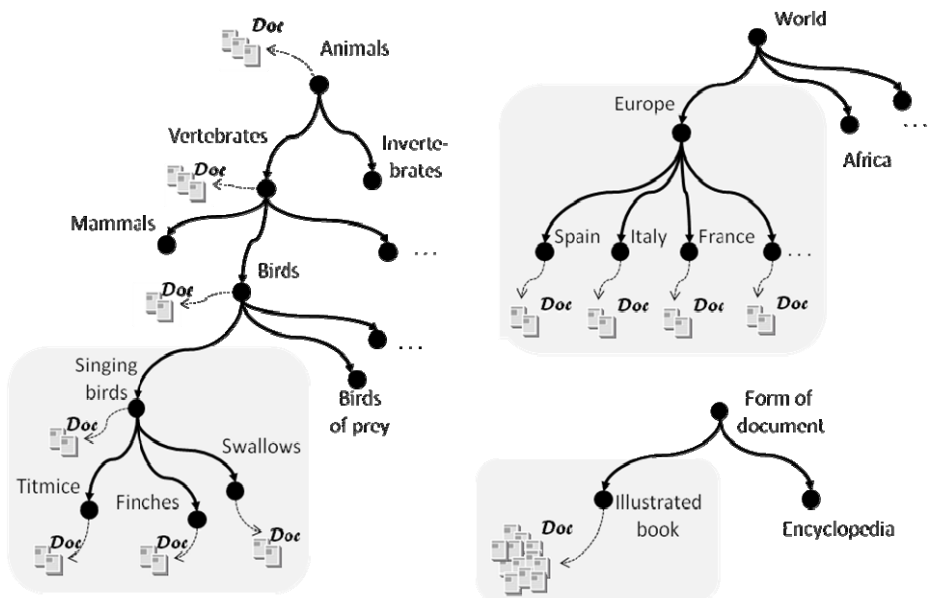

Fig. 3: Knowledge structure with general facets, hierarchical inferences and associated documents

If the entities of the knowledge representation are organized by aspects, the production of thematic sections can be done for each aspect separately. Thus, an aspect-oriented search processes is prepared. Figure 3 shows a simple example.

If this approach should work over several stages of the relational paths, a suitable preparation of the corresponding hierarchies is required. This requirement can be best met if no polyhierarchies are established. This especially means that there is a strict adherence to the criterion that each hierarchically subordinated concept has all the characteristics of the parent concept. Then it is assured that the inheritance of characteristics can take place over several stages of the relational paths. The formal requirement for such an inheritance is described by the property of *transitivity*. Transitivity as a formal property of a relation ~ means that from *a* ~ *b* and *b* ~ *c* follows that *a* ~ *c* is valid.

Transferred to an understanding of hierarchy between conceptual entities this means that from the two statements

>entity B is a narrower concept to entity A

and

>entity C is a narrower concept to entity B

it can be formally concluded:

>entity C is a narrower concept to entity A.

It should be emphasized that this property belongs to the relation and does not have to be checked at conceptual properties of the entities A, B and C. If any relation has the property of transitivity, then it applies to all entities that are connected by the relation within the knowledge representation. This describes an important difference in the consideration of indexing and formal knowledge representation languages. In the context of indexing languages relationships between entities are generally interpreted by analyzing the properties of the entities content-wise. Neither they are reported formally as a characteristic property nor are they seen as a general property of the relationships. In indexing languages one frequently meets the situation that the transitivity between hierarchically related concepts sometimes is present and sometimes it is not. For transferring the potential of reasoning along relational paths from formal knowledge representations to information retrieval, this situation is not acceptable. Instead, the appropriate formal requirements for the definition of entities and relationships must be respected. The aim must be to transfer content heritability of desired characteristics of entities to the corresponding formal relationships between entities. On this basis, general inference processes along the relational paths can be designed to be used as beneficial retrieval tools.

The property of inheritance of characteristics has obvious advantages for the formal design of concept structures. The figure 4 shows a simple example of a concept structure in which any property has been assigned to each concept explicitly.

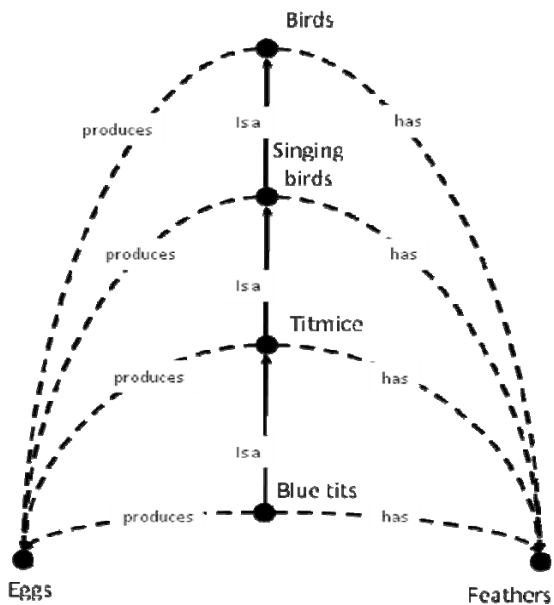

Fig. 4: Knowledge structure with redundant property declarations

Figure 5 shows the same concept structure as a formal representation with the feature of inheritance of characteristics, illustrated for only one of the properties.

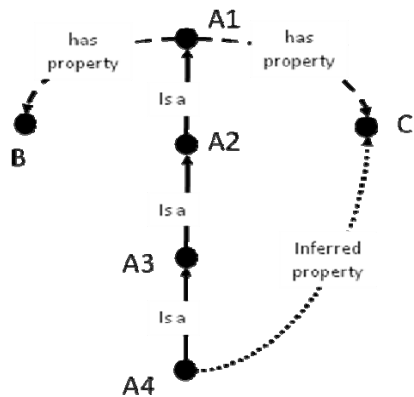

Fig. 5: Knowledge structure with inferred properties

The representation can be read as follows:

$A_n$ is a $A_{n-1}$

$A_1$ has property C

=> $A_n$ has property C

This is consistent with the objective of a knowledge representation to make the represented knowledge formally interpretable and to derive conclusions about the relational paths. From the design of expert systems it is well-known that this approach has its limits - to be seen for example in figure 4 when considering *penguin* as an additional concept - which must be treated by exception clauses. The compatibility between cognitive interpretation and assignment of formal attributes remains a specific challenge for the design of knowledge representation languages to be used for the purposes of indexing and information retrieval. Our approach avoids the necessity to specify exception clauses.

*2.2 Typed relations in knowledge representations*

A possible way to improve the formal properties of relationships in indexing languages is described by an extension of the relational spectrum that is taken as the basis for the relationships modeled between the entities of a knowledge representation. The starting point of our discussion will be the three types of relationships that are known from the classical indexing languages: synonymy, hierarchy, and association relationship. Particularly interesting is which statements can be made about the possibilities of reasoning, if relationships of the same or of different types are combined to a joint relational path. Let us first summarize the relationships in question with their respective properties.

*Synonymy or equivalence relationship*

Formal knowledge representations do not distinguish between different denominations of an entity. Their remit is not based on indexing purposes. Therefore, no distinction between indexing and access vocabulary is needed. The semantic content of an entity is expressed by any of their linguistic denominations. However, the distinction between indexing and access vocabulary is a traditional and important component of indexing languages and should be retained as part of our considerations. This relationship type, however, need not be specifically included in the considerations for transitivity, since it can occur only as the first relation of a path and the properties of the resulting path depend on the characteristics of the other relationships involved.

*Hierarchy relationships*

The concept of hierarchy relations allows for the differentiation in several types of relationships. The *generic relationship* is the model case, which is connected with the strict definitional default that a narrower concept possesses *all* characteristics of the generic term as well as at least a further. Thought along but frequently unexpressed with this is, that these additional features must originate from a common aspect area or a certain categorical facet, we will call this also a *generic context*. Only the strict attention of this condition provides a problem-free combination of an idea of super- and sub-ordination of concepts with the formal property of transitivity. If the generic context for the determination of features is changed, polyhierarchies become unavoidable and the result most commonly is a pre-combined conceptual ordering by insufficiently expressed criteria. As an example

> table, wooden table, glass table, kitchen table, living room table, art nouveau table, desk, changing table, side table.

Any attempt to organize these concepts - or even an extension - into a single hierarchy must fail because of the reasons stated. Therefore, a subordination by changing aspects, as it is often found in pre-combined classifications or in thesauri with extensive use of conceptual composition (e. g. by using compound nouns), cannot be a suitable condition for drawing conclusions by relational inferences. To support inferences we need a faceted structure preserving the generic contexts within each hierarchy string. This may mean that even the generic relationship must be differentiated into subtypes according to an appropriation of the knowledge structure to be modeled (Boteram, 2008).

The second type of hierarchy relationships is described by the *whole-part relationship*. A part can never have all the properties of the associated whole, therefore no transitivity can be expected over several stages for this type of relationship. However, the formation of relational paths in combination with other relationships has to be considered. For these considerations, the sequence of the path formation will play an important role.

A linear order is established also by the chronological relationship, in form of a *later-earlier-* or a *earlier-later-relationship* it is part of certain indexing languages. If periods are formed, a hierarchical relationship can be derived, which in case of the same direction of the time arrow is usable for transitive conclusions.

*Associative relationships*

The third group of relationships in indexing languages is given by the so-called *associative relationships*. Typically represented in thesauri by one single abbreviation (commonly *RT - Related Term*), this abbreviation can usually identify several different types, which are not specified for simplicity of use. The RT-relationship is intended to support retrieval operations by referring to conceptual related terms. Their origin can be seen as a result of declaring synonyms for an indexing language. Many of the terms of such similarity clusters can be characterized as *near synonyms*, terms with similar but no identical conceptual meaning. These clusters must be resolved for the conceptual modeling of indexing languages by declaring appropriate terms as synonyms or alternatively as related terms. Decisions about the existence of an association between concepts are at best supported by considerations of consistency criteria, but not by formal specifications. Therefore, generally no transitivity can be expected along the paths of the association relationship used in thesauri. Analyzes in existing indexing languages confirm this expectation.

For giving an example we choose the *ASIST thesaurus* with its focus on information science and technology (ASIS&T Thesaurus, 2005). The initial entity should be the descriptor *Automatic indexing*. Figure 6 shows descriptors that are associated to *Automatic indexing* as related terms by varying path length.

| **Automatic indexing** | | |
| --- | --- | --- |
| **RT of path length 1** | **RT of path length 2** | **RT of path length 3** |
| Automatic classification | Automatic categorization | Categories |
| Computational linguistics | Cluster analysis | Classification |
| Content based indexing | Computational lexicography | Co-occurence analysis |
| Information processing | Full text searching | Cognitive science |
| Machine aided indexing | Image indexing | Computer science |
| Natural language processing | Information science | Cybernetics |
| | Knowledge representation | Data presentation |
| | Natural language interfaces | Domain analysis |
| | Probabilistic indexing | Image analysis |
| | Relevance ranking | Image databases |
| | Text processing | Image retrieval |
| | | Images |
| | | Information retrieval |
| | | Information science education |
| | | Information scientists |
| | | Information technology |
| | | Librarianship |
| | | Linguistics |
| | | Ontologies |
| | | . . . |

Fig. 6: Paths of associative relations from the *ASIST thesaurus*

A trail of associative terms beginning with *Automatic indexing* and ending with *Images* or *Cognitive science* cannot be seen as suitable for creating thematic clusters with a sharp conceptual border.

A number of studies suggest strong benefit for indexing and retrieval purposes by elaborating a finer structure of the associative relationships (Tudhope, Alani & Jones, 2001; Green & Bean, 2006; Bean & Green, 2001; Michel, 1997). With our attempt of linking the conceptual interpretation of the entities with the benefits of formal inference processes along the relationship paths, we want to continue these proposals. For this purpose we have condensed the current proposals for specifying associative relationships furthermore and placed the result in a list that is shown in figure 7.

- Raw material / product
- Causality (cause – effect)
- Action/ product
- Person as actor / action
- Institution as actor / action
- Person as actor / product
- Institution as actor / product
- Field of application or reference

Fig. 7: Inventory of typed associative relations

Using this inventory, the results of our analyzes can be summarized in tabular overviews. The first table (figure 8) contains transitivity statements when connecting relationships of the same type.

| Type of relation | Relation 1 | Relation 2 | Transitivity |
|---|---|---|---|
| Equivalence | Synonym | Synonym | O |
| Hierarchy | Abstraction, generic context | Abstraction, generic context | + |
|  | Whole / Part | Whole / Part | + |
|  | Abstraction, generic context | Whole / Part | - |
|  | Whole / Part | Abstraction, generic context | - |
| Chronological context | Earlier / Later | Earlier / Later | + |
|  | Later / Earlier | Later / Earlier | + |
|  | Earlier / Later | Later / Earlier | - |
|  | Later / Earlier | Earlier / Later | - |
| Association | Unspecific assoziation | Unspecific assoziation | - |
|  | Raw material / product | Raw material / product | + |
|  | Causality (cause – effect) | Causality (cause – effect) | + |
|  | Person as actor / action | Person as actor / action | - |
|  | Institution as actor / action | Institution as actor / action | - |
|  | Person as actor / product | Person as actor / product | - |
|  | Institution as actor / product | Institution as actor / product | - |
|  | Action/ product | Action/ product | - |

+: Transitivity is given
-: Transitivity cannot be expected
O: Not allowed for indexing languages

Fig. 8: Transitivity in case of same type of relationships

Most statements about transitivity are self-explanatory, the others can be explained by the aid of simple examples. For the synonym relationship, transitivity in indexing languages cannot be studied, since a synonym can never be both source and destination for a relationship. The labeling was carried out in figure 8.

The table of figure 9 contains statements about transitivity when connecting hierarchical and chronological relationships of different types. A statement for connecting synonyms with other relationships is repeated, synonyms as referral forms can never appear as the target of a relationship.

| Type of relation | Relation 1 | Relation 2 | Transitivity |
|---|---|---|---|
| Hierarchy | Synonym | Abstraction, generic context | + |
| | Synonym | Whole / Part | + |
| | Abstraction, generic context | Synonym | O |
| | Abstraction, generic context | Synonym | O |
| | Whole / Part | Synonym | O |
| | Whole / Part | Synonym | O |
| | | | |
| Chronological context | Synonym | Earlier / Later | + |
| | Synonym | Later / Earlier | + |
| | Earlier / Later | Synonym | O |
| | Later / Earlier | Synonym | O |
| | Abstraction, generic context | Earlier / Later | + |
| | Abstraction, generic context | Later / Earlier | + |
| | Earlier / Later | Abstraction, generic context | + |
| | Later / Earlier | Abstraction, generic context | + |
| | Whole / Part | Earlier / Later | + |
| | Whole / Part | Later / Earlier | + |
| | Earlier / Later | Whole / Part | + |
| | Later / Earlier | Whole / Part | + |
| +: Transitivity is given<br>-: Transitivity cannot be expected<br>O: Not allowed for indexing languages | | | |

Fig. 9: Transitivity in case of different type of hierarchical relationships

The transitions from hierarchical to associative structures are represented in a model-like manner in the relational graphs of figure 10 and figure 12. It was already mentioned that for statements about the transitivity of the transition from one structure to another, it is important to observe the order of the connections between the relations. Figure 10 shows the transition from an associative to a hierarchical structure.

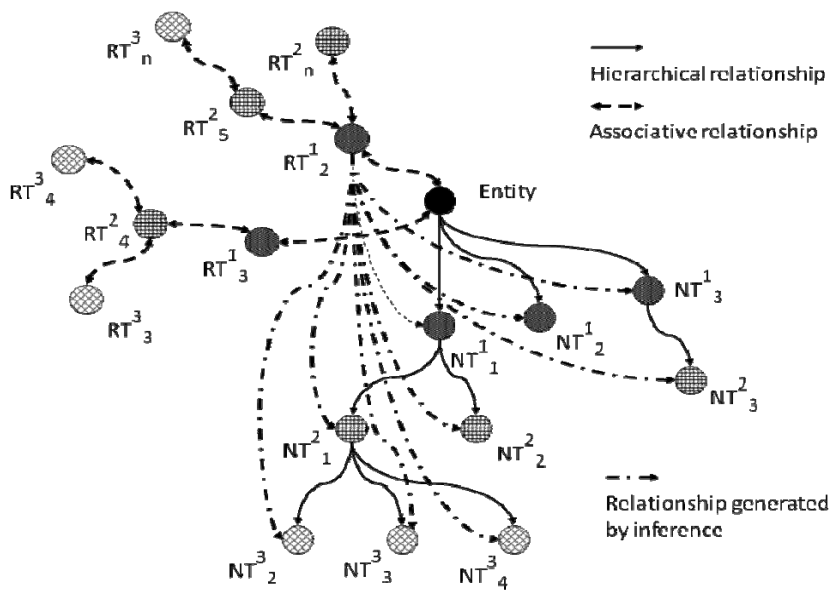

Fig. 10: Transition from associative relationships into a hierarchical structure

An interpretation of this situation reads as follows. A transition from a related term to another can only be justified by its benefit for a content-oriented retrieval process. Therefore, any evaluation is bound to criteria that are beyond the possibilities of formalization within the concept structure. A continuation into a hierarchical structure organized by aspect differentiation, which in turn is characterized by transitivity between its stages, therefore is generally possible. While the first part of the continuation can only be justified by content-oriented arguments, the second part can be evaluated by formal characteristics. Figure 11 lists the transitive connections of different types of associative relations as first relationship with hierarchical relations as second relationship in form of a table.

| Type of relation | Relation 1 | Relation 2 | Transitivity |
|---|---|---|---|
| Association | Unspecific assoziation | Abstraction, generic context | + |
| | Unspecific assoziation | Whole / Part | + |
| | Unspecific assoziation | Earlier / Later* | + |
| | Raw material / product | Abstraction, generic context | + |
| | Raw material / product | Whole / Part | + |
| | Raw material / product | Earlier / Later* | + |
| | Action/ product | Abstraction, generic context | + |
| | Action/ product | Whole / Part | + |
| | Action/ product | Earlier / Later* | + |
| | Person as actor / action | Abstraction, generic context | + |
| | Person as actor / action | Whole / Part | + |
| | Person as actor / action | Earlier / Later* | + |
| | Institution as actor / action | Abstraction, generic context | + |
| | Institution as actor / action | Whole / Part | + |
| | Institution as actor / action | Earlier / Later* | + |
| | Causality (cause – effect) | Abstraction, generic context | + |
| | Causality (cause – effect) | Whole / Part | + |
| | Causality (cause – effect) | Earlier / Later* | + |
| | Person as actor / product | Abstraction, generic context | + |
| | Person as actor / product | Whole / Part | + |
| | Person as actor / product | Earlier / Later* | + |
| | Institution as actor / product | Abstraction, generic context | + |
| | Institution as actor / product | Whole / Part | + |
| | Institution as actor / product | Earlier / Later* | + |
| * Also: Later / Earlier | | | |

Fig. 11: Transitivity for combinations of typed associative with hierarchical relationships

Figure 12 shows the transition from a hierarchical to an associative structure. A formally determined transitive continuation cannot be expected for this case in general. Reasons have already been mentioned.

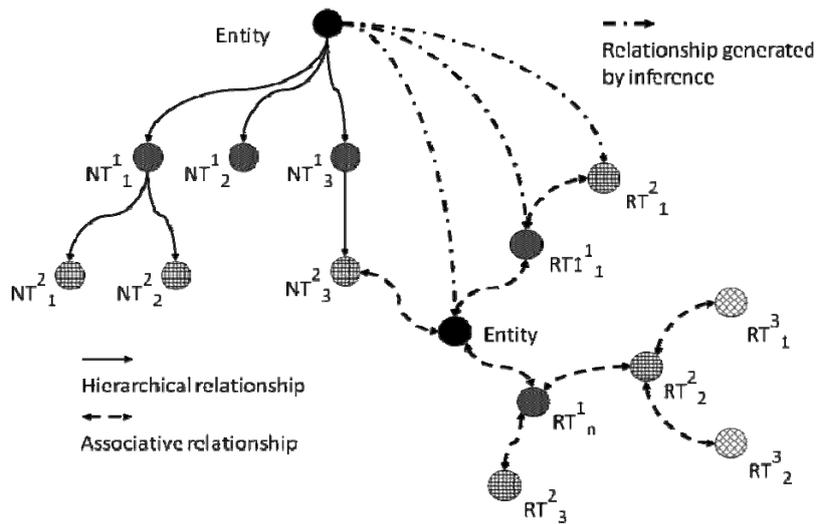

Fig. 12: Transitions from a hierarchical structure to associative relationships

We do not want to specify the non-transitive combinations of different types of associative relationships.

As a result of the analyzes we can note, that typing of the associative relationships offer more possibilities for formal inference procedures along the relational paths of a conceptual structure, as it can be expected from the usual context of indexing languages. If one is interested in the continuation of the path outlined, it remains a task for further investigation to determine a suitable inventory of typed relations with their formal properties.

We now want to discuss what benefits the combination of typed relations and inferences may have for retrieval operations.

*2.3 Typed relations and inferences in information retrieval*

Our model for ontology-based indexing and retrieval consists of a combination of elements of traditional indexing and formal knowledge representation. The entities as conceptual representatives should be ordered into facets. Within the facets the conceptual structure should be limited to transitive hierarchical relationships. Between the entities of different facets typed associative relationships are established, whose formal properties have to be characterized in detail for inference processes. The formal properties of the knowledge representation support the formulation of conditions for selecting the entities. The established relationships are used for this purpose. Inference processes can help to generate sets of hits without all terms of the query formulation must have been indexed.

As an illustration of the model, we consider once more the examples indicated at the outset of the paper

- We are looking for documents about the *migratory instinct <u>of</u> songbirds*,
- We are looking for documents about *songbirds <u>with</u> migratory instinct*.

It has been noted that the traditional indexing and retrieval model offers no solution for the second problem. We now want to clarify which type of modeling is enabled by our approach and what contribution such a modeling can provide for the solution of our problem.

A simultaneous indexing with e.g. the entities *singing birds* and *migration pattern* cannot be applicable for the second topic. Our approach allows considering the knowledge modeling in figure 13 first, before choosing any terms for performing a search. It uses typed relations apart from a hierarchically arranged taxonomy of the bird species to specify additional properties of the modeled entities. This modeling is also not suitable for our problem because the statement *singing birds have migration pattern* is in fact incorrect, as is the statement *singing birds eat grains*. The incorrectness arises from the interpretation as a statement about *all* songbirds and the non-existent ability to specify the statement to the appropriate species only. There are songbirds with migratory instinct, but also those without, and there are songbirds that eat grains, but also those that do not.

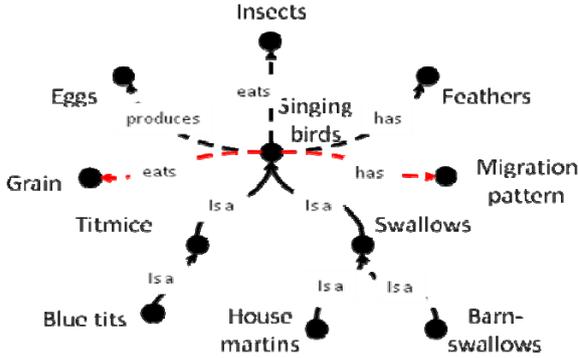

Fig. 13: Knowledge structure with incorrect connection of typed relations

In order to avoid that the specification of properties by typed relations leads to incorrect statements, the entities may be connected together only at certain appropriate points in the hierarchy. As a rule, it can be said that the statements of the typed relations may not be assigned completely to each tree of a hierarchical taxonomy (single mammalian species also have a migratory instinct), but that the modeling must be done selectively in accordance with the principles of aspect-oriented faceting. For our example, a *behaviour facet* in addition to a *taxonomy facet* would have to be considered (2). If the necessary relations are established for each entity specifically, a considerable effort is required. Furthermore, the result may be a web of relationships that holds great potential for lack of clarity in itself. The figure 14 gives a small impression for our example.

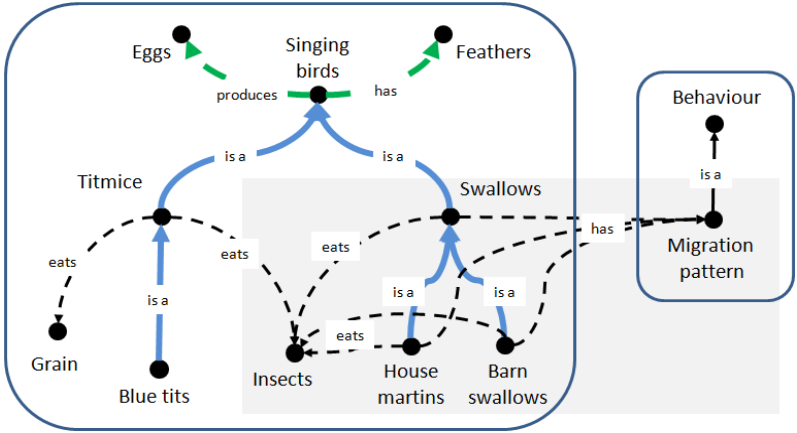

Fig. 14: Knowledge structure with typed property assignments but without inferences

Drawing inferences along the paths of the hierarchical structures offer the possibility to counteract this complexity. If drawing inferences is possible, a relationship that is established by a typed relation is not only valid between the entities directly involved, but it is transferred formally to all respective terms. The content-orientated consistency of the formal properties has to be ensured by choosing the correct entities for establishing the connection. For our example this means, one has to select the suitable entities of the highest hierarchical level. This solution is indicated in figure 15.

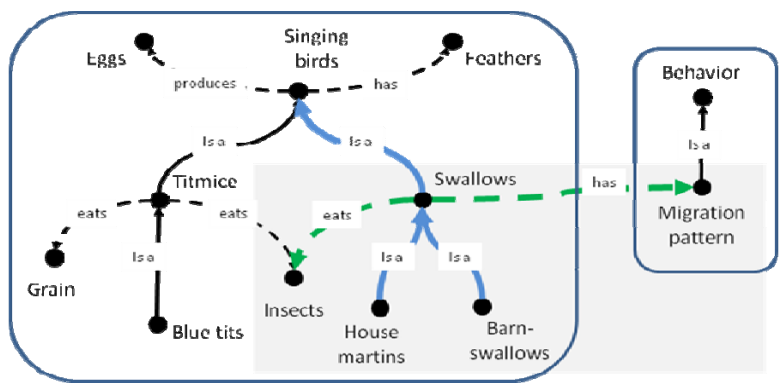

Fig. 15: Knowledge representation with typed property assignments and with inferences

By applying these design options we can indicate a solution to our question: We are looking for documents about *songbirds <u>with</u> migratory instinct*. For this answer neither *songbirds* nor *migratory instinct* must have been indexed as entities to the documents. We are looking for documents that make statements *about* certain entities (*songbirds*) with specified properties (*have migratory instinct*) without these properties themselves must be treated content. The taxonomy facet of our knowledge representation contains the knowledge about songbird species. A typed relation provides the connection to the property *has migratory instinct* for the appropriate entities. An inference about the hierarchical relation paths allows to form the set of all applicable songbirds and thus to carry out a respective search. In contrary, an answer to our first question - We are looking for documents about the *migratory instinct <u>of</u> songbirds* - would require that *migratory instinct* can be searched as indexed entity as well as *songbirds*. The retrieval facility must allow for a post-coordinated combination of the entities. We can summarize that the combination of using typed relations with formal inferences permits the solution of a problem that cannot be solved by classical methods alone.

## 3   A prototype

In order to demonstrate the opportunities discussed here within an interactively searchable information system, an experimental prototype was developed at the Institute of Information Science at the University of Applied Sciences Cologne (3). Some of its components and options to illustrate our results will be presented here. The prototype was not constructed as an ideal-typical system, instead existing data were used. These data consist of approximately 14,000 data records that were taken from a database *Literatur zur Informationserschließung* (4). The scope of the documents corresponds to an extract from the *ASIST Thesaurus* (ASIS&T, 2005), which was transformed into a *topic map* (ISO 13250, 1999; Pepper; Garshol, 2005; Park & Hunting, 2003) by using the software *Ontopia* (5). Already existing relationships between the descriptors were analyzed and enhanced with typed relations according to an extended relation inventory. *Ontopia* uses the *XTM* format as a formal representation environment. The formal specification of *XTM* (Pepper & Moore, 2001) can be used for specifying and enhancing the set of relationships. As an example we present the declaration for distinguishing the roles *isMethodOf* (role 1) and *isAdopting* (role 2) of the typed associative relation *Methodology*:

```xml
<!-- Präambel 1-A: Typed associative relations -->
<!-- Relation -->
<topic id="Methodology">
  <baseName>
    <instanceOf>
      <topicRef xlink:href="#AssociativeRelation"/>
    </instanceOf>
    <baseNameString>Associative Relation (Methodology)</baseNameString>
  </baseName>
</topic>

<!-- role 1 -->

<topic id="isMethodOf">
  <baseName>
    <instanceOf>
      <topicRef xlink:href="#MethodologyRelationMember"/>
    </instanceOf>
    <baseNameString>is method of</baseNameString>
  </baseName>
</topic>

<!-- role 2 -->

<topic id="isAdopting">
  <baseName>
    <instanceOf>
      <topicRef xlink:href="#MethodologyRelationMember"/>
    </instanceOf>
    <baseNameString>is adopting</baseNameString>
  </baseName>
</topic>
```

The connection between two entities - *abstracting_and_indexing_service_bureaus* and *indexing* - by the relation *Methodology* with the pair of roles *isAdopting* and *isMethodOf* then looks as follows:

```xml
<!-- Präambel 2: Definition of entity roles of typed relations -->
  <association>
    <instanceOf>
      <topicRef xlink:href="#Methodology"/>
    </instanceOf>
    <member>
      <roleSpec>
        <topicRef xlink:href="#isAdopting"/>
      </roleSpec>
      <topicRef xlink:href="#abstracting_and_indexing_service_bureaus"/>
    </member>
    <member>
      <roleSpec>
        <topicRef xlink:href="#isMethodOf"/>
      </roleSpec>
      <topicRef xlink:href="#indexing"/>
    </member>
  </association>
```

Figure 16 shows an excerpt of our topic map for the entity *index languages* with its typed relations to other entities.

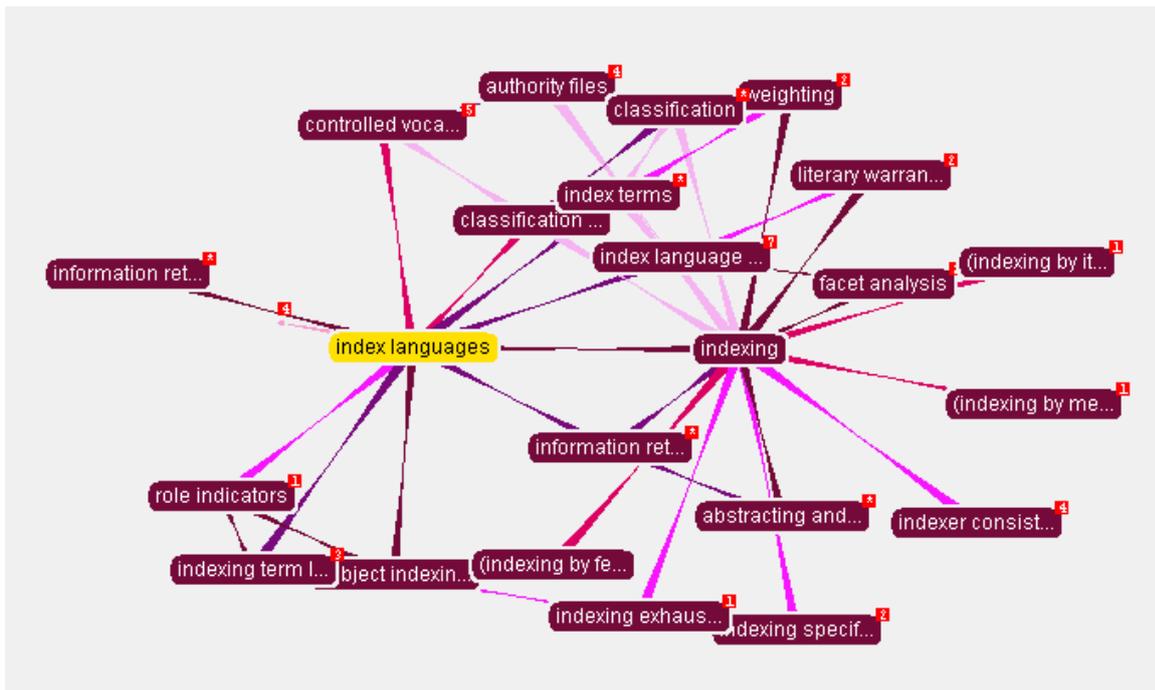

Fig. 16: *Ontopia* topic map for an excerpt from the *ASIST Thesaurus* with additional typed relations

Using a Web-interface (6) and the *Ontopia* specific query language *tolog* (Garshol; Tolog, 2007) allows querying the names of the topics and also using the modeled relationships between the topics as selection criteria. Figure 17 shows the search template with a sample query.

Fig. 17: Web interface for searching the *ASIST* topic map and corresponding documents

The request with *tolog* displays the resulting topics. These topics are the basis for generating the set of documents that appear as a result of the search.

For the evaluation of the results it should be considered that the indexing of the 14,000 documents with descriptors of the *ASIST thesaurus* was based on an automatic indexing procedure using the software *Lingo* (Lepsky & Vorhauer, 2006) (7). The subject indexing quality thus may be suboptimal compared with an intellectual indexing.

We give two examples of querying for different relationships of the entity *indexing* and discuss them in some more detail. Figure 17 shows a request for documents that considers the entity *indexing* as *methodic procedure*. The selection of the topics as *tolog* input (left side of figure 17) reads:

```
Methodology($TOPIC : isMethodOf, indexing : isAdopting)?
```

The related topics are displayed in figure 17 below the heading *Topics* (upper right side). For our search query we receive four topics:

```
facet analysis
index languages
literary warrant
weighting
```

This set of topics corresponds to the approach using typed relations in our *ASIST* topic map for the topic *Indexing*, namely the relation *Methodology*, as is shown in figure 18:

Relations (5):

- Associative Relation (Characterization)
    - is characterizing
        - indexer consistency
        - indexing exhaustivity
        - indexing specificity
- Associative Relation (Methodology)
    - is adopting
        - facet analysis
        - index languages
        - literary warrant
        - weighting
    - is method of
        - abstracting and indexing service bureaus
- Associative Relation (Production)
    - is producing
        - information retrieval indexes
- Associative Relation (Usage)
    - is using
        - authority files
        - classification
        - classification schemes
        - index terms
- Hierarchical Relation
    - Broader Term
        - (indexing by feature indexed)
        - (indexing by item indexed)
        - (indexing by method used)

Fig.18: Typed relations for the topic *indexing* modeled in the *ASIST* topic map

Under the heading *Documents* in figure 17, twelve documents are displayed that have been indexed with one of the four topics (lower right side).

Our second example asks for documents about *tools* or *instruments* of the topic *indexing*. The *tolog* input with the topic *Indexing* and the relation *Usage* reads:

```
Usage($TOPIC : isInstrumentOf, indexing : isUsing)?
```

The identified topics are in this case:

```
authority files
classification
classification schemes
index terms
```

They correspond to the modeling of the relation *Usage* for the topic *Indexing* (cf. the box with dashed lines in figure 18). The determined document set in this case consists of 565 documents.

As a conclusion it can be stated that the use of typed relations for knowledge modeling allows to select different sets of topics for differentiated searches. Thus, different sets of hits can be formed, each of which contains the documents for the specified search interests.

Our prototype also allows to combine a selection of typed relations with inferences about the hierarchical structure of the topic map. For a search on the topic *Controlled_vocabularies*, we will include first all narrower terms. In this case, a hierarchical expansion has to be performed and the query needs to be supplemented by inference rules. Within the *tolog* context, such inference rules are called *Custom Inference Rules*. Our query interface provides a separate input box for this purpose, which is shown in figure 19 down left.

Fig. 19: Search-interface for the *ASIST* Topic map with *Custom Inference Rules*

The following rules are predefined. They can be activated by the link "*sample*". If desired they can be substituted with own rules.

```
direct-narrower-term($A, $B) :-
  HierarchicalRelation($A : broaderTermMember,
                      $B : narrowerTermMember).

strictly-narrower-term($A, $B) :- {
  direct-narrower-term($A, $B) |
  direct-narrower-term($A, $C), strictly-narrower-term($C, $B)
}.

narrower-term($A, $B) :- {
  $A = $B | strictly-narrower-term($A, $B)
}.

narrower-term-1($A, $B) :- {
  $A = $B | direct-narrower-term($A, $B)
}.

narrower-term-2($A, $B) :- {
  narrower-term-1($A, $B) |
```

```
    narrower-term-1($A, $C), narrower-term-1($C, $B)
  }.

  narrower-term-3($A, $B) :- {
    narrower-term-2($A, $B) |
    narrower-term-2($A, $C), narrower-term-1($C, $B)
  }.

  direct-broader-term($A, $B) :-
    direct-narrower-term($B, $A).

  strictly-broader-term($A, $B) :-
    strictly-narrower-term($B, $A).

  broader-term($A, $B) :-
    narrower-term($B, $A).

  broader-term-1($A, $B) :-
    narrower-term-1($B, $A).

  broader-term-2($A, $B) :-
    narrower-term-2($B, $A).

  broader-term-3($A, $B) :-
    narrower-term-3($B, $A).
```

Topics should be searched that represent *Requirements* in the sense of the modeled relation *Production* for producing *Controlled_vocabularies* regardless of their special type. Within the topic map this corresponds to the modeling by *Controlled_vocabularies* as a topic and the direction *isProducing* of the relation *Production*. The type of indexing language should not be distinguished, which gives cause to the hierarchical inference.

The *tolog* input for the query reads:

```
Production($TOPIC : isProducing, $PRODUCT : isProductOf),
narrower-term(controlled_vocabularies, $PRODUCT)?
```

Five topics are found that are associated with 687 documents:

```
automatic indexing (96)
index language construction (0)
subject heading lists (45)
subject headings (530)
vocabulary control (41)
```

The second query

```
Production($TOPIC : isProductOf, $PRODUCT : isProducing),
narrower-term(controlled_vocabularies, $PRODUCT)?
```

is used for all products that are produced with the help of *controlled_vocabularies* (including all narrower concepts). This corresponds to the modeling by *controlled_vocabularies* as a topic and the direction *isProductOf* of the relation *production* and the additional inclusion of the hierarchical inference. Two topics are found with 502 associated documents:

```
authority files (140)
thesauri (372)
```

Other examples can be processed by using the provided search interface.

## 4 Assessment of the approach and outlook

Our discussion has shown how an ontology based model for indexing and retrieval may look like that uses typed relations for refined representation of semantic knowledge and inference processes for executing search processes. The prototype shows the distribution of tasks between indexing and selection of search terms.

Concluding our discussion, we want to draw attention to a further aspect, for which only provisional results can be submitted. Maybe these initial results stimulate to further research.

With the combination of faceted structures and the provision of typed relations between entities of different facets, it is possible to express issues that are commonly expressed only by aids of syntactic indexing. We want to indicate the interrelationships.

We have chosen typed relations that are typical elements of *phase relations* or *Citation orders*, as developed for example in the context of the *Colon Classification*, the *Bliss Classification, 2nd ed.*, or the indexing system *PRECIS* (Ranganathan, 1987; BC2, 1992; Austin, 1969a; Austin, 1969b; Austin, 1984).

This raises the question, whether the procedure described can represent and make searchable issues that otherwise afford the use of syntax elements for expressing the a posteriori context in the document. Figure 20 shows a prototypical situation of a faceted knowledge structure, symbolized as a systematic arrangement.

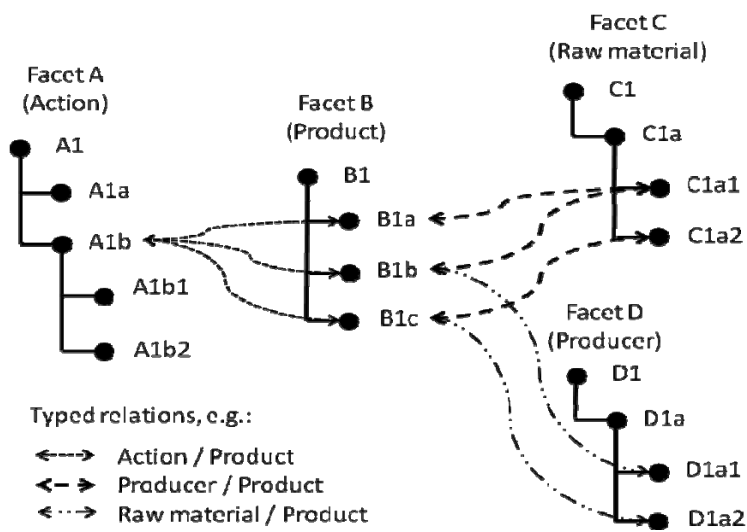

Fig. 20: Faceted systematic structure with typed relations between the facets

By syntactic indexing a document-specific relationship between entities of different facets is expressed through a synthesized index string. The sequence of the elements contained in this string is formed according to a citation order, we take for an example the synthetic notation D1a1B1bA1b. In our model, this correlation would be represented via the connections, which are modeled by typed relations in the knowledge structure. The complete flexibility of syntactic indexing for free synthesis of document-specific issues is not attainable by this approach. Depending on the established relational structure, a variety of topics can be expressed that are not specifically representable by the means of coordinate indexing and Boolean retrieval. Thus, the previously proposed approaches for the use of faceted structures (Gödert, 1991; Broughton, 2001; Tunkelang, 2009) in retrieval environments can be enriched by new ideas. However, we need to emphasize the exemplary character of our instances. For the moment, they should only illustrate the methodological approach. More precise statements about the potential and the qualitative properties of the procedure require further research, including the development of appropriate test environments and scenarios.

**Footnotes**

(1) From the multitude of presentations for designing indexing languages and their relationship types we will cite only the standard ISO 25964 (ISO 25964, 2011-2013).
(2) A suitable explanation of the underlying principles for the production of faceted classification structures may be found in a book by Brian Buchanan (Buchanan, 1979).
(3) Special thanks to *Matthias Nagelschmidt* and *Jens Wille* who set up the contentual and technical prerequisites of the search environment and thus allows performing the first experiments as well as verifying our statements.
(4) Cf.: http://ixtrieve.fh-koeln.de/LitIE/.
(5) Cf.: http://www.ontopia.net.
(6) Cf.: http://ixtrieve.fh-koeln.de/ghn.
(7) This software is especially powerful in the identification of multi-word groups and their matching with a pre-defined dictionary. Cf.: http://lex-lingo.blogspot.com.
(8) The search can be performed by using the interface: http://ixtrieve.fh-koeln.de/ghn/.

Author:   Prof. Winfried Gödert
          Cologne University of Applied Sciences
          Institute of Information Science
          Gustav-Heinemann-Ufer 54
          D-50968 Köln
          winfried.goedert@fh-koeln.de